\documentclass[conference]{IEEEtran}
\usepackage{CJK}
\usepackage{amsmath}
\usepackage{amsfonts}
\usepackage{amssymb}
\usepackage{bm}
\usepackage{subeqnarray}
\usepackage{algorithm}
\usepackage{algorithmic}
\usepackage{cases}
\usepackage{ tipa }
\usepackage{graphicx}
\usepackage{epstopdf}
\usepackage{float}
\usepackage{verbatim}
\usepackage{color,xcolor}
\usepackage{makecell}
\begin{document}
\title{RL-Based User Association and Resource Allocation for Multi-UAV enabled MEC }

\author{\IEEEauthorblockN{Liang Wang\IEEEauthorrefmark{1}, Peiqiu Huang\IEEEauthorrefmark{2}, Kezhi Wang\IEEEauthorrefmark{1}, Guopeng Zhang\IEEEauthorrefmark{3}, Lei Zhang\IEEEauthorrefmark{4}, Nauman Aslam\IEEEauthorrefmark{1} and Kun Yang\IEEEauthorrefmark{5}}	
	\IEEEauthorblockA{\IEEEauthorrefmark{1}Department of Computer \& Information Sciences, Northumbria University, Newcastle upon Tyne, UK}
	\IEEEauthorblockA{\IEEEauthorrefmark{2}School of Informacion Science \& Engineering, Central South University, China}
	\IEEEauthorblockA{\IEEEauthorrefmark{3}School of Computer Science \& Technology, China University of Mining and Technology, China}
\IEEEauthorblockA{\IEEEauthorrefmark{4}School of Engineering, University of Glasgow, UK}	
	\IEEEauthorblockA{\IEEEauthorrefmark{5}University of Electronic Science and Techology of China, Zhongshan Institute, China}	
	\IEEEauthorblockA{\IEEEauthorrefmark{5}School of Computer Science \& Electronic Engineering, University of Essex, UK}

}

\maketitle
\begin{abstract}
In this paper, multi-unmanned aerial vehicle (UAV) enabled mobile edge computing (MEC), i.e., UAVE is studied, where several UAVs are deployed as flying MEC platform to provide computing resource to ground user equipments (UEs). Compared to the traditional fixed location MEC, UAV enabled MEC (i.e., UAVE) is particular useful in case of temporary events, emergency situations and on-demand services, due to its high flexibility, low cost and easy deployment features. However, operation of UAVE faces several challenges, two of which are how to achieve both 1) the association between multiple UEs and UAVs and 2) the resource allocation from UAVs to UEs, while minimizing the energy consumption for all the UEs. To address this, we formulate the above problem into a mixed integer nonlinear programming (MINLP), which is difficult to be solved in general, especially in the large-scale scenario. We then propose a Reinforcement Learning (RL)-based user Association and resource Allocation (RLAA) algorithm to tackle this problem efficiently and effectively. Numerical results show that the proposed RLAA can achieve the optimal performance with comparison to the exhaustive search in small scale, and have considerable performance gain over other typical algorithms in large-scale cases.

\end{abstract}

\begin{IEEEkeywords}
Reinforcement Learning, Mobile Edge Computing, Unmanned Aerial Vehicle, User Association, Resource Allocation
\end{IEEEkeywords}

\IEEEpeerreviewmaketitle

\section{Introduction}
Nowadays, user equipments (UEs) such as smart phones, tablets, wearable devices and other Internet of smart things are becoming increasingly popular and bringing huge convenience to our daily life. Moreover, many emerging mobile applications (e.g., augmented reality, smart navigation and interactive service) are receiving more and more attention but most of those applications are resource intensive, which makes the UEs very difficult to execute them,
due to limited battery and computation resource (e.g. CPU, storage or memory) in UEs. 

Fortunately, mobile edge computing (MEC) has recently been proposed as a means to enable UEs with intensive computational tasks to offload them to the edge cloud, which can not only prolong the battery life of UEs, but also increase UEs' computational capacity. Offloading decision making and resource allocation have been studied in~\cite{8274943, 4427231}, while MEC with Cloud Radio Access Network (C-RAN) has been investigated in~\cite{8304391,8037986,7590428}. The above works either consider there is only one MEC (e.g., \cite{8274943, 6787113}), or consider the MECs have fixed location (e.g.,~\cite{8353131,8304391}), which may not be practical in some scenarios. For instance, the single MEC is normally resource-limited and may not be able to meet the requirement of all the UEs at the same time. Also, MEC with fixed location lacks flexibility and may not be suitable to the cases where the number and the requirement of UEs keep changing.
  
Unmanned aerial vehicle (UAV), due to the features of  low cost, high flexibility and easy to deployment, have recently attracted much attention in wireless communication, e.g., serving as base station~\cite{7762053} or mobile relays~\cite{8424236}. UAV enabled MEC (e.g.,~\cite{8438896}) have been proposed by integrating MEC server to UAVs (i.e., UAVE), to provide computing resource to ground UEs. Compared with the traditional fixed location MEC, UAVE is of particular interest to the scenario such as 1) temporary events (i.e., in case of a large number of people gathering in the ground celebrating a big event or watching football match); 2) emergency situations (i.e., in case of earthquake and the infrastructure may be destroyed or temporary unavailable) or other on-demand services. However, the operation of UAVE faces many challenges, two of which are how to achieve 1) the association between multiple UEs and UAVs and 2) the resource allocation from the UAVs to the UEs, while meeting the quality of service (QoS) and minimizing the whole energy consumption for all the UEs.

To address these challenges, we formulate above problem into a mixed integer non-linear programming (MINLP), which is very difficult to be addressed in general, especially in the large-scale scenario (e.g., when there is a large number of UEs in the ground waiting to be served). We then propose a Reinforcement Learning-based user Association and resource Allocation (RLAA) algorithm to deal with this problem efficiently and effectively. Numerical results show that the proposed RLAA can achieve the optimal performance compared to the exhaustive search in small scale, and have considerable performance gain over other typical algorithms in large-scale scenarios.

The rest of the paper is organized as follows. We show the system model and the optimization problem in Section \uppercase\expandafter{\romannumeral2}. Then, our proposed RLAA algorithm is introduced in Section \uppercase\expandafter{\romannumeral3}. The simulation result is given in Section \uppercase\expandafter{\romannumeral4}, followed by the conclusion remarks in Section \uppercase\expandafter{\romannumeral5}.

\section{System Model}
\begin{figure}[t]
	\centering
	\includegraphics[width=3.5in]{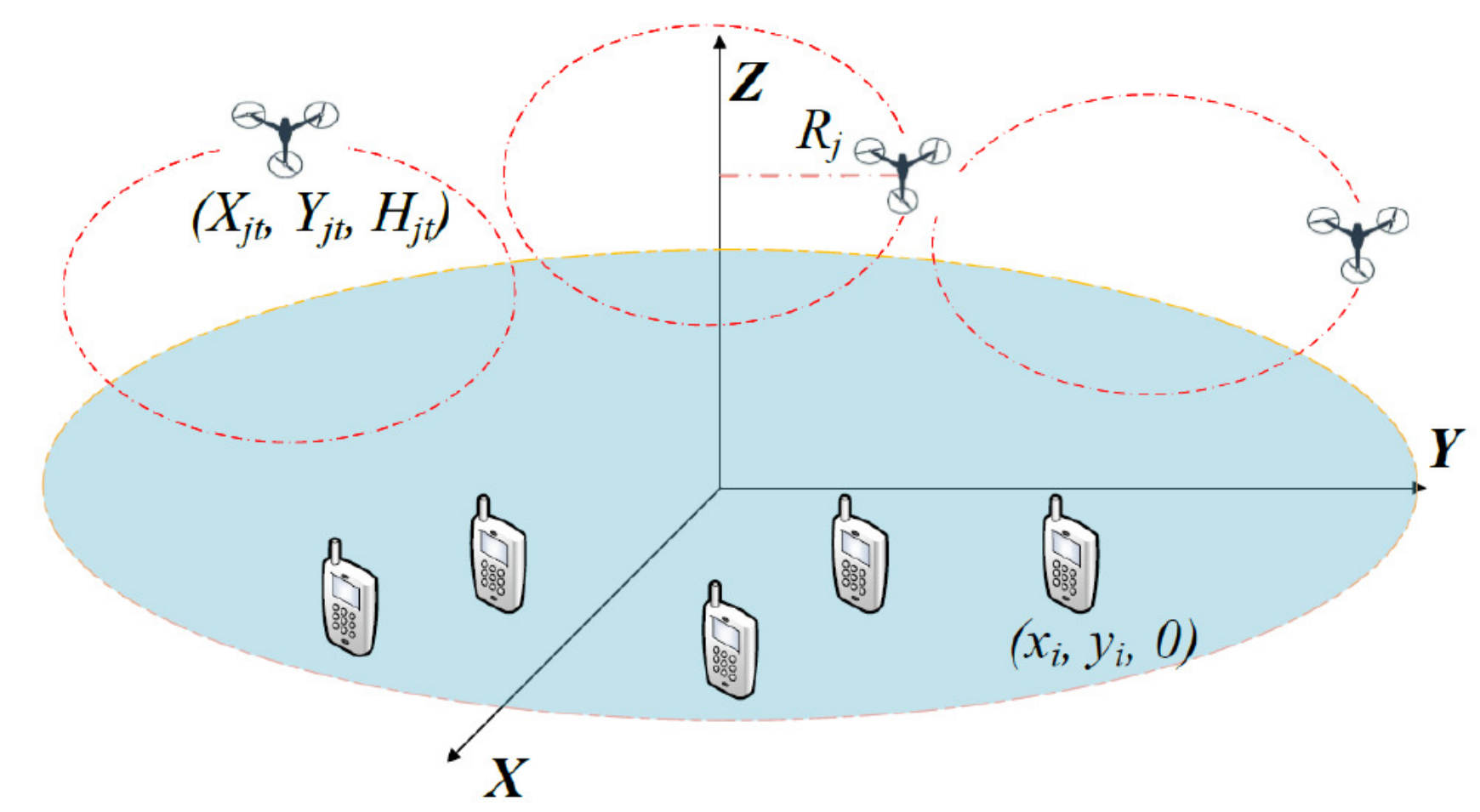}
	\caption{A Multi-UAV enabled MEC system	
	}\label{fig1}
\end{figure}

As shown in Fig.~\ref{fig1}, we consider there are $i \in \mathcal{N}=\{1, 2,..., N\}$ UEs, each of which has a computation-intensive task to be executed.
Also, we consider there are $j \in \mathcal{M}=\{1, 2,..., M\}$ UAVs deployed as the MEC platform, flying in a circle with radius $R_j$. 
Define a new vector $j \in  \mathcal{M'}=\{0, \mathcal{M}\}$ to denote the possible place where the tasks from ground UEs can be executed at, in which $j=0$ denotes that UE conducts task itself without offloading. 
Similar to \cite{8438896}, we assume that the $j$-th UAV's flight period can be discretized into $t \in\mathcal{T}_j=\{1, 2,..., T_j\}$ time slots. Define a new vector $t \in  \mathcal{T}'_j=\{0, \mathcal{T}_j\}$ to denote the possible time slots when the tasks from ground UEs can be executed at, in which $t=0$ denotes that UE conducts task itself. Also we assume that the UAV's location change within each time slot can be ignored, compared to the distances from the UAV to all UEs.

Denote the coordinate of the $j$-th UAV at $t$-th time slot as
$[X_{jt}, Y_{jt}, H_{jt}]$ and the coordinate of the $i$-th UE as $[x_i,y_i,0]$.

Similar to \cite{8304391}, assume $i$-th UE has a computational intensive task ${I}_{i}$ to be executed as
\begin{equation}\label{w1}
{I}_{i}=({D}_{i}, {F}_{i}),~ \forall i \in \mathcal{N}
\end{equation}
where ${F}_{i}$ denotes the total number of CPU cycles required to complete this task and ${D}_{i}$ denotes the amount of data needed to be transmitted to UAV if deciding to offload, in which ${D}_{i}$ and ${F}_{i}$ can be obtained by using the approaches provided in~\cite{6253581}. Assume that each UE can decide either to execute the task locally or choose to offload to one of the UAVs in one time slot and also assume that the task can be completed in this time slot. Similar to \cite{8274943}, we do not consider the time for returning the results back to UE from UAV. 
Thus, one can have 
\begin{equation}\label{w2}
\begin{aligned}
C1: {a}_{ijt}\text{ = \{0,1\}, }\forall i\in \mathcal{N}, j\in \mathcal{M'}, t \in \mathcal{T}'_j
\end{aligned}
\end{equation}
where ${a}_{ijt}= 1$, $j \neq 0$, $t \neq 0$ denotes that the $i$-th UE choose the $j$-th UAV in the $t$-th time slot to offload, while ${a}_{ijt}= 1$, $j = 0$, $t=0$ denotes that $i$-th UE execute the task itself and otherwise, ${a}_{ijt}= 0$. Note that $t=0$, if and only if $j=0$.

Also, assume that the $j$-th ($j \in \mathcal{M}$) UAV can serve more than one UE in each time slot and this task has to be completed either via offloading or local execution. Therefore, one can have
\begin{equation}\label{w3}
\begin{aligned}
C2: \sum_{j=0}^{M}\sum_{t=0}^{T_j}a_{ijt}=1,~ \forall i \in \mathcal{N}
\end{aligned}
\end{equation}

\subsection{Task Offloading}
In offloading scenario, we assume the horizontal distance between $i$-th UE and the $j$-th UAV in $t$-th time slot as
\begin{equation}\label{w4}
\begin{aligned}
R_{ijt}=\sqrt{(X_{jt}-x_{i})^2+(Y_{jt}-y_{i})^2}
\end{aligned}
\end{equation}
Then, the offloading data rate can be given by
\begin{equation}\label{w5}
\begin{aligned}
r_{ijt}=B\log_2(1+\frac{\alpha P_{i}^{Tr}}{H_{jt}^2+R_{ijt}^2}), \forall i \in \mathcal{N}, j \in \mathcal{M}, t \in \mathcal{T}_j
\end{aligned}
\end{equation}
where $B$ is denoted as the channel bandwidth, $P_{i}^{Tr}$ as the transmission power of the $i$-th UE, $\alpha$=$\frac{g_0G_0}{\sigma^2}$, $G_0$ $\approx$ \text{2.2846}, $g_0$ as the channel power gain at the reference distance \text{1} $m$ and $\sigma^2$ as the noise power~\cite{8103781}. 

Also, one can see that the time to offload the data from  $i$-th UE to the $j$-th UAV in $t$-th time interval can be given as
\begin{equation}\label{w6}
{T}_{ijt}^{Tr}=\frac{{D}_{i}}{{r}_{ijt}},~\forall i\in \mathcal{N}, j \in \mathcal{M}, t \in \mathcal{T}_j
\end{equation}
Also, the time to execute the task can be expressed as  
\begin{equation}\label{w7}
\begin{aligned}
T_{ijt}^{C}= \frac{F_i}{f_{ijt}},~ \forall i \in \mathcal{N}, j \in \mathcal{M}, t \in \mathcal{T}_j
\end{aligned}
\end{equation}
where $f_{ijt}$ is the computation resource that the $j$th UAV could provide to the $i$-th UE. Then, we can have the total time consumption as
\begin{equation}\label{w8}
\begin{aligned}
T_{ijt}= T^{Tr}_{ijt}+T^C_{ijt},~ \forall i \in \mathcal{N}, j \in \mathcal{M}, t \in \mathcal{T}_j
\end{aligned}
\end{equation}
Moreover, the total energy consumption of the $i$-th UE to the $j$-th UAV in $t$-th time slot can be given as
\begin{equation}\label{w9}
\begin{aligned}
E^{Tr}_{ijt}=P_{i}^{Tr}T^{Tr}_{ijt},~ \forall i \in \mathcal{N}, j \in \mathcal{M}, t \in \mathcal{T}_j
\end{aligned}
\end{equation}

Similar to \cite{8274943}, we assume each UAV in every time slot can only accept limited amount of offloaded task. Then, one has
\begin{equation}\label{w10}
\begin{aligned}
C3: \sum_{i=1}^{N}a_{ijt} \leq K,~ \forall j \in \mathcal{M}, t \in \mathcal{T}_j
\end{aligned}
\end{equation}
where $K$ is the maximal number of UEs that each UAV can accept in each time slot.

\subsection{Local Execution}
If the UE decides to execute the task locally, the power consumption for the $i$-th UE can be given by $k_i(f_{ijt})^{v_i}$, where $j=0$, $t=0$, and $k_i$ $\geq$ \text{0} is the effective switched capacitance and $v_i$ can be normally to \text{3} \cite{8274943}. Then, the local execution time can be given by 
$
T^C_{ijt}=\frac{F_i}{f_{ijt}}
$, ($j=0$, $t=0$) and then, the total energy consumption can be given as
$k_i{(f_{ijt})}^{v_i} T_{ijt}^C$.

\subsection{Problem Formulation}
Then, one can have the energy consumption of each UE as
\begin{equation}\label{w11}
\begin{aligned}[l]
E_{ijt} = 
\begin{cases}
k_i {(f_{ijt})}^{v_i} T^C_{ijt}, & \text{if $j=0$, $t=0$} \\
P^{Tr}_{i}T^{Tr}_{ijt}, & \text{if } j \in \mathcal{M}, t \in \mathcal{T}_j \\
\end{cases}
\end{aligned}
\end{equation}
Also, the total time spent to complete each task can be expressed as
\begin{equation}\label{w12}
\begin{aligned}
T_{ijt}=
\begin{cases}
T^C_{ijt}, & \text{if $j=0$, $t=0$} \\
T^{Tr}_{ijt}+T^C_{ijt}, &\text{if } j \in \mathcal{M}, t \in \mathcal{T}_j \\
\end{cases}
\end{aligned}
\end{equation}	
One can assume that the maximal computation resource which the $j$-th UAV can provide is as $f_{j}^{max}$. Then, one can have
\begin{equation}\label{w13}
\begin{aligned}
C4: \sum_{i=1}^{N}a_{ijt}f_{ijt} \leq f_{j}^{max},~ \forall j \in \mathcal{M}, t \in \mathcal{T}_j
\end{aligned}
\end{equation}
Also, as the task normally has to be completed in certain amount of the time and thus without loss of generality, we assume the task must be completed in time $T^{max}$ without loss of generality.
In our paper, assume all the transmitting and computing process for each task must be completed within one time interval $T^{max}$, Then, we have
\begin{equation}\label{w14}
\begin{aligned}
C5: \sum_{j=0}^{M}a_{ijt}T_{ijt} \leq T^{max},~ \forall i \in \mathcal{N}, t \in \mathcal{T}'_j
\end{aligned}
\end{equation}


Denote $\bm{A}$ = $\{a_{ijt}, \forall i \in \mathcal{N}, j \in \mathcal{M'}, t \in \mathcal{T}'_j\}$, $\bm{F}$ = $\{f_{ijt}, \forall i \in \mathcal{N}, j \in \mathcal{M'}, t \in \mathcal{T}'_j\}$. Then, one can have

\begin{subequations}\label{w15}
\begin{align}
\mathcal{P}: & \underset{\bm{A},\bm{F}}{\text{min}}\sum_{i=1}^{N}\sum_{j=0}^{M}\sum_{t=1}^{T_j}a_{ijt}E_{ijt} \\
& \text{subject to }\\
&C1: a_{ijt} \in \{0,1\},~\forall i\in \mathcal{N}, j \in \mathcal{M'}, t \in \mathcal{T}'_j\\
&C2: \sum_{j=0}^{M}\sum_{t=0}^{T_j}a_{ijt}=1, \forall i \in \mathcal{N}\\
&C3: \sum_{i=1}^{N}a_{ijt} \leq K,~ \forall j \in \mathcal{M}, t \in \mathcal{T}_j\\
&C4: \sum_{i=1}^{N}a_{ij}f_{ijt} \leq f_{j}^{max},~ \forall j \in \mathcal{M}, t \in \mathcal{T}_j\\
&C5: \sum_{j=0}^{M}a_{ijt}T_{ijt} \leq T^{max} \text{, } \forall i \in \mathcal{N}, t \in \mathcal{T}'_j
\end{align}
\end{subequations}

Note that the above problem $\mathcal{P}$ is a MINLP problem, which is difficult to be solved optimally in general. Some existing algorithms like exhaustive search or branch and bound algorithm may solve this problem, but with prohibitive complexity. Therefore, in this paper, we aim to obtain an efficient solution to solve this problem. To this end, we propose the RLAA algorithm to deal with $\mathcal{P}$ effectively and efficiently.

\section{Proposed Algorithm}
In this section,  we show our proposed RLAA algorithm. First, we introduce three important elements in RLAA (i.e., actions, states, and reward functions).
\begin{itemize}
	\item \textbf{Actions:} At each episode $eps$, each UE takes an action. If the UE decides to offload the task to the $j$-th UAV in $t$-th time interval, the action is denoted as $\rho_{jt}$, $\forall  j \in \mathcal{M}, t \in \mathcal{T}_j$. If UE decides to execute the task locally, the action is as $\rho_{00}$. Then, one can define the collection of actions as follows:
	\begin{equation}
	\mathcal{C} =\{\rho_{00},\rho_{11},...,\rho_{1T_1},...,\rho_{M1},...,\rho_{MT_M}\}.
	\end{equation}
	
	\quad For above offloading action $\rho_{jt}$, $\forall j \in \mathcal{M}, t \in \mathcal{T}_j$, the minimal computation resources of the $i$-th UE can be given by
	\begin{equation}\label{omcr}
	f_{ijt}^{min}=\frac{F_i}{T^{max}-\frac{D_i}{r_{ijt}}},~ \forall j \in \mathcal{M}, t \in \mathcal{T}_j
	\end{equation}
	For local execution action $\rho_{00}$, the minimal computation resources of the $i$-th UE is given as
	\begin{equation}\label{lmcr}
	f_{ijt}^{min}=\frac{F_i}{T^{max}},~j=0, t=0
	\end{equation}
	
	\quad Note that not all actions can guarantee that the task can be completed within one time interval, as the available computation resources may be less than the minimal computation resources (i.e., $f_{ijt}^{min}$ in~\eqref{omcr} and~\eqref{lmcr}). Similarly, the communication resource can also not be guaranteed (i.e., $\mathcal{C}3$ in (15e)). Therefore we may remove some actions in $\mathcal{C}$, resulting in the collection of feasible actions for the $i$-th UE as $\mathcal{C}_i$. 
	\item \textbf{States:} Then, we define the states as follows:
	\begin{equation}
	{\bf{s}}=\{\omega_{1},\ldots,\omega_{i},\ldots,\omega_{N}\}
	\end{equation}
	where $\omega_{i}$ represents the decision of the $i$-th UE.  Specifically, if the $i$-th UE $(i\in \mathcal{N})$ offloads the task to the $j$-th UAV in $t$-th time interval, we assign action $\rho_{jt}$ $\forall j \in \mathcal{M}, t \in \mathcal{T}_j$ to state $\omega_{i}$. It is worth mentioning that if the $i$-th UE decides to execute the task locally, we assign action $\rho_{00}$ to state $\omega_{i}$.
	\item \textbf{Reward Functions:} We define the reward function as 
	\begin{equation}\label{or}
	Z({\bf{s}},\rho_{jt})=\frac{1}{E_{ijt}}	
	\end{equation}
	\quad The above proposed reward function can keep reducing the energy consumption of each UE and may finally achieve the minimization of the energy consumption of all UEs.
\end{itemize}

Then, we present RLAA in Algorithm~\ref{RLAA}. In the beginning, states $\bf{s}$ is initialized. The $Q$-table is also initialized, which is used to record every state and action (i.e., line 1 in Algorithm~\ref{RLAA}). At each episode, we obtain the collection of the actions $\mathcal{C}_i$ for the $i$-th UE. Then, according to the $\epsilon$-greedy policy~\cite{mnih2015human}, the $i$-th UE either chooses a random action with probability $\epsilon$ or follows the greedy policy with probability $1-\epsilon$, which is expresses as
\begin{equation}\label{policy}
\begin{aligned}
\rho_{jt}=
\begin{cases}
\rho, & \text{if~} \text{rand}(0,1)<\epsilon \\
\underset{\rho_{jt} \in \mathcal{C}_i}{\text{argmax}}Q({\bf{s}},\rho_{jt}), & \text{otherwise} \\
\end{cases}
\end{aligned}
\end{equation}
where $\rho$ is an action randomly selected from $\mathcal{C}_i$, rand(0,1) denotes a random number uniformly distributed over the interval [0,1] (i.e., line 4 - line 8 in Algorithm~\ref{RLAA}).

Then, the resource allocation is conducted for the $i$-th UE (i.e., line 9 in Algorithm~\ref{RLAA}). If the $i$-th UE offload the task to the $j$-th UAV in $t$-th time slot, the minimal computation resource $f_{ijt}^{min}$ in ~\eqref{omcr} is allocated. If the $i$-th UE execute task locally, the minimal computation resource $f_{ijt}^{min}$ in ~\eqref{lmcr} is allocated. Based on the proposed reward function in~\eqref{or}, the $i$-th UE can then obtain a reward (i.e., line 10 in Algorithm~\ref{RLAA}).


Next, we update the $Q$-table (line 11), where the updating rule of $Q$-table is given as 
\begin{equation}\label{w23}
\begin{aligned}
Q({\bf{s}},c) \leftarrow &Q({\bf{s}},c)+\beta\{Z({\bf{s}},c)\\&+\gamma\underset{c \in \mathcal{C}_i}{\text{max}}Q({\bf{s}'},c)-Q({\bf{s}},c)\},\\
\end{aligned} 
\end{equation}
where $\gamma$ is the reward decay over the interval [0,1], $\beta$ is the learning rate  over the interval [0,1], and ${\bf{s}'}$ is the next state. Also, states ${\bf{s}}$ is updated based on action $\rho_{jt}$. Specifically, we assign action $\rho_{jt}$, $\forall j \in \mathcal{M}', t \in \mathcal{T}'_j$  to state $\omega_{i}$. 

The above process will be repeated until the maximum episode ($eps^{max}$) is reached. Finally, each UE selects an action according to $Q$-table (line 16). Specifically, for the $i$-th UE, the action in $C_i$ corresponding to the largest value of $Q$-table is selected.

\begin{algorithm}[htbp]
	\caption{Our proposed RLAA}\label{RLAA}
	\begin{algorithmic}[1]
		\STATE Initialize $\bf{s}$ and $Q$-table;
		\STATE $\textbf{while}$ ${eps}$ $\leq$ ${eps}^{max}$ 
		\STATE ~~$\textbf{for}$ $i$ = 1:$N$
		\STATE ~~~~$\textbf{if}$ \text{rand}(0,1)$<\epsilon$
		\STATE ~~~~~~Select an action $\rho_{jt}$ randomly from $\mathcal{C}_i$ for the $i$-th UE;
		\STATE ~~~~$\textbf{else}$ 
		\STATE ~~~~~~Select an action $\rho_{jt}=\underset{\rho_{jt} \in \mathcal{C}_i}{\text{argmax}}Q({\bf{s}},\rho_{jt})$ for the $i$-th UE;
		\STATE ~~~~$\textbf{end if}$\
		\STATE ~~~~~Allocate computation resource $f_{ijt}^{min}$ from~\eqref{omcr} and~\eqref{lmcr} for the $i$-th UE;
		\STATE ~~~~Obtain a reward $Z({\bf{s}},\rho_{jt})$ according to~\eqref{or};
		\STATE ~~~~Update $Q$-table according to~\eqref{w23};
		\STATE ~~~~Update $\bf{s}$;
		\STATE ~~$\textbf{end for}$
		\STATE ~~$\text{$eps$} \leftarrow \text{$eps$}+1$;
		\STATE  $\textbf{end while}$
		\STATE  Select an action for each UE.
	\end{algorithmic}
\end{algorithm}

\section{Simulation Results}

\begin{table}[!t]
	\centering
	\caption{Simulation Parameters}
	\begin{tabular}{|c|c|c|c|}
		\hline
		\textbf{Parameters} & \textbf{Settings} \\ \hline
		\text{Radius $R_j$ for all UAVs} & \text{800 m}   \\ \hline
		\text{Flying height $H_{jt}$ for all UAVs} & \text{350 m} \\ \hline
		\text{Bandwidth $B$ } & \text{1 MHz} \\ \hline
		\text{Transmission power $P_i^{Tr}$} & \text{1 W} \\ \hline
		\text{Noise variance $\sigma^2$} & \text{$-90$ dbm/Hz} \\ \hline
		\text{$G_0$} & \text{2.2846} \\ \hline
		\text{Channel power gain $g_0$} & \text{1.42}$\times 10^{-4}$ \\ \hline
		\text{Data Size $D_i$} & \text{[$100, 1000$] KB} \\ \hline
		\text{Execution task $F_i$} & \text{[$10^8, 10^9$] cycles} \\ \hline
		\text{Time duration $T^{max}$} & \text{1 s} \\ \hline
		\text{Location of UEs} & \text{$[-2000, 2000] \times [-1000, 1000]$ m} \\ \hline
		\text{$f_j^{max}$} & \text{150} GHz \\ \hline
		\text{$\epsilon$-greedy policy probability} & \text{$0.9$} \\ \hline
		\text{Reward decay $\gamma$} & \text{0.9} \\ \hline
		\text{Learning rate $\beta$} & \text{0.2} \\ \hline
		\text{$k_i$ for all UEs} & \text{$10^{-27}$} \\ \hline
		\text{$v_i$ for all UEs} & \text{3} \\ \hline
		\text{$eps^{max}$} & \text{10000} \\ \hline
		\text{$T_j$ for all UAVs} & \text{12} \\ \hline
	\end{tabular}
\label{tab1}
\end{table}
In this section, the simulation for the proposed multi-UAV enabled MEC system is conducted, where the parameters of the tests are shown in Table.~\ref{tab1}, in which the channel bandwidth is set to $B$ = 1 MHz, the noise variance is set to $\sigma^{2}$ = $-90$ dbm/Hz, the channel power gain at the reference distance 1 $m$ is set to $g_0$ = $1.42 \times 10^{-4}$~\cite{8103781}, the transmission power $P_i^{Tr}$ is set to 1 W, the time interval $T^{max}$ is set to 1 s, the $k_i$ is set to $10^{-27}$ for all the UEs. Also, we assume each UAV can support $K$ = 150 UEs in one time slot. All UEs are assumed to be randomly distributed in a rectangle area of coordinates $[-2000, 2000] \times [-1000, 1000]$ m. We randomly select the data size $D_i$ of each task from the interval of $[100, 1000]$ KB and select $F_i$ from the interval of $[10^8, 10^9]$ cycles. 

In order to evaluate the performance of our proposed RLAA, the following four algorithms are used as comparison algorithms.
\begin{itemize}
	\item \textbf{Exhaustive search (ES):} We examine all the possibilities, with the objective of minimizing the overall energy consumption for all the UEs.
	\item \textbf{Local execution (LE):} We assume all tasks are executed locally and there are no offloading.
	\item \textbf{Random offloading (RO):} Each UE randomly selects the UAV and the time slot to offload its task.
	\item \textbf{Greedy offloading (GO):} Each UE selects the nearest UAV to offload its task. If the UAV is overloaded (i.e., $C3$ is violated), then selects the second nearest UAV to offload and so on. 
\end{itemize} 

\begin{figure}[!t]
	\centering
	\includegraphics[width=3.5in]{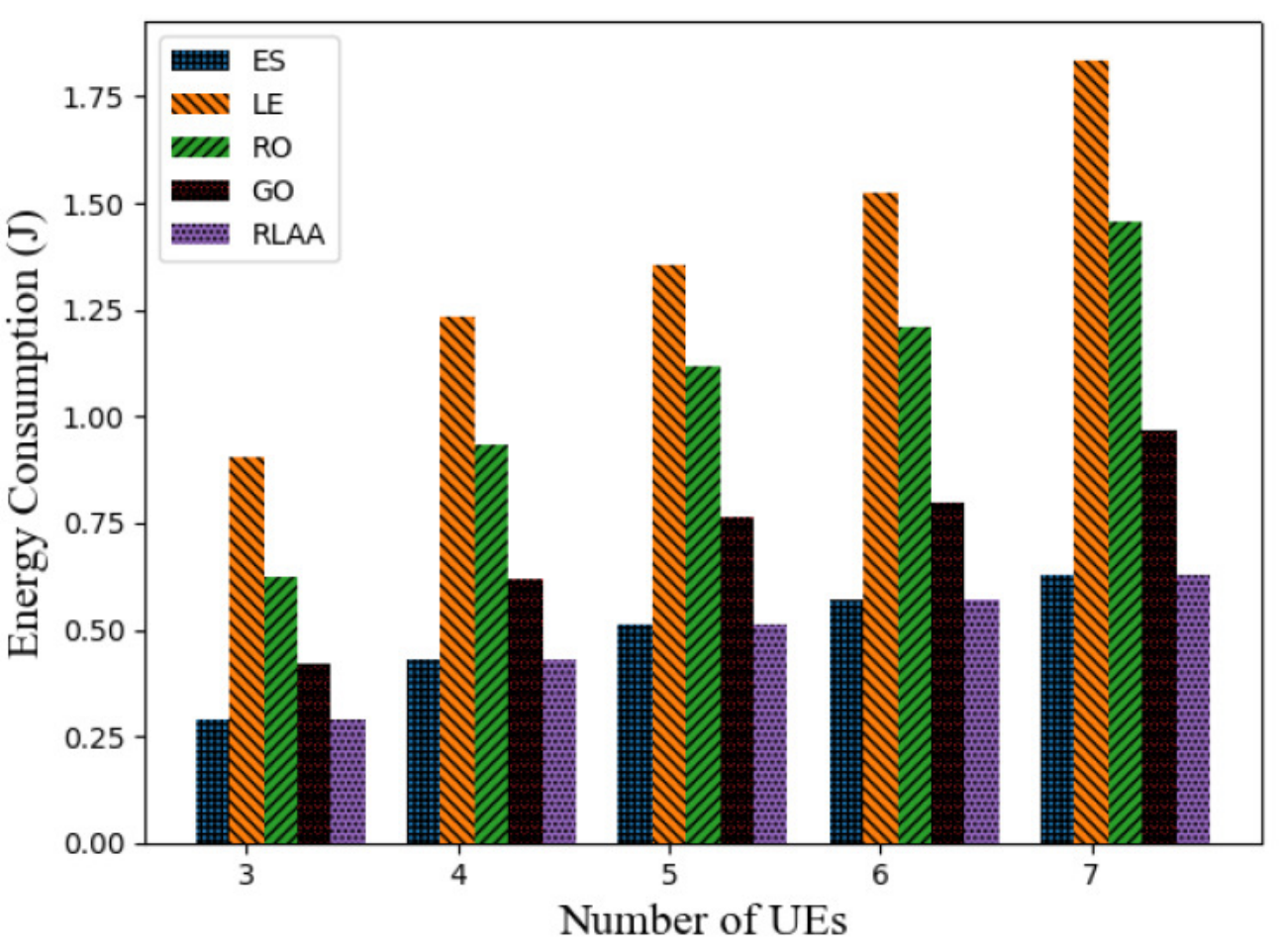}
	\caption{The overall energy consumption of ES, LE, RO, GO and RLAA versus the number of UEs.}\label{fig2}
\end{figure}

Firstly, we compare the performance of RLAA with its four compared algorithms on a set of small scale instances (i.e., the number of UEs ranges from 3 to 7). We assume that there are two UAVs flying in circles with the same radius and the center coordinates of two UAVs are set to $[1200, 1200, 350]$ and $[-1200, -1200, 350]$, respectively. From Fig.~\ref{fig2}, one can see that RLAA has the same performance as ES, both of which can achieve the minimal enery consumption. Also, one can see that GO achieves better performance than RO, whereas LE achieve the worst performance for all examined values. This is because that our proposed RLAA can choose most energy-efficient action for all the UEs according to computation and communication requirement, while others either make UE to execute all the task locally (i.e., LE), or randomly offload the tasks (i.e., RO), or just find the nearest UAV (i.e., GO), resulting in worse performance.  

\begin{figure}[!t]
	\centering
	\includegraphics[width=3.5in]{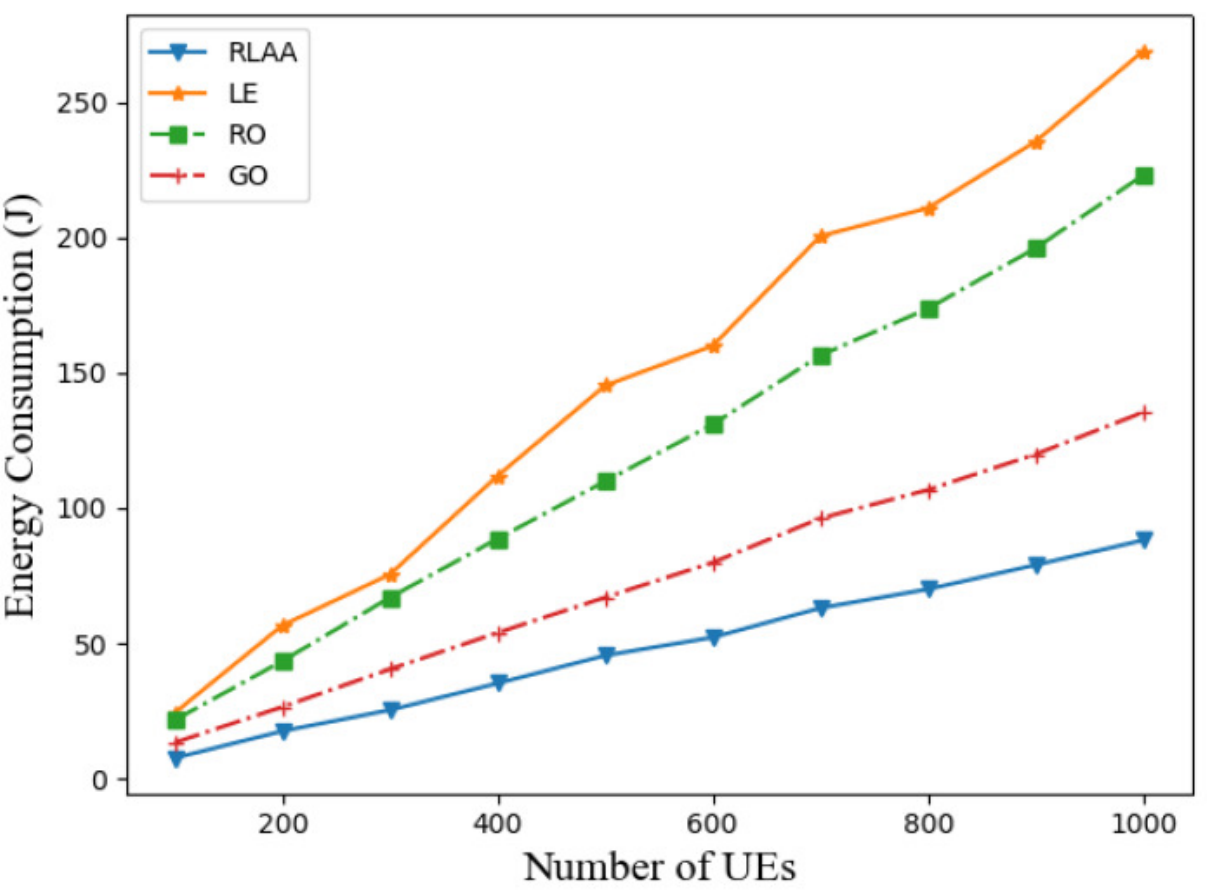}
	\caption{The overall energy consumption of RLAA, LE, RO and GO versus the number of UEs with 3 UAVs.}\label{fig3}
\end{figure}

Next, we compare the performance of RLAA with LE, RO and GO on a set of large scale instances, where the number of UEs is increased to 100$\sim$1000.
The number of the UAVs is set to 3, where the center coordinates are $[1200,1200,350]$, $[-1200,-1200,350]$ and $[-1200,1200,350]$, respectively. Note that we do not examine ES here, due to its prohibitive complexity. From Fig.~\ref{fig3}, one can see that our proposed RLAA still performs best, followed by GO, RO and LE, as expected.   

\begin{figure}[!t]
	\centering
	\includegraphics[width=3.5in]{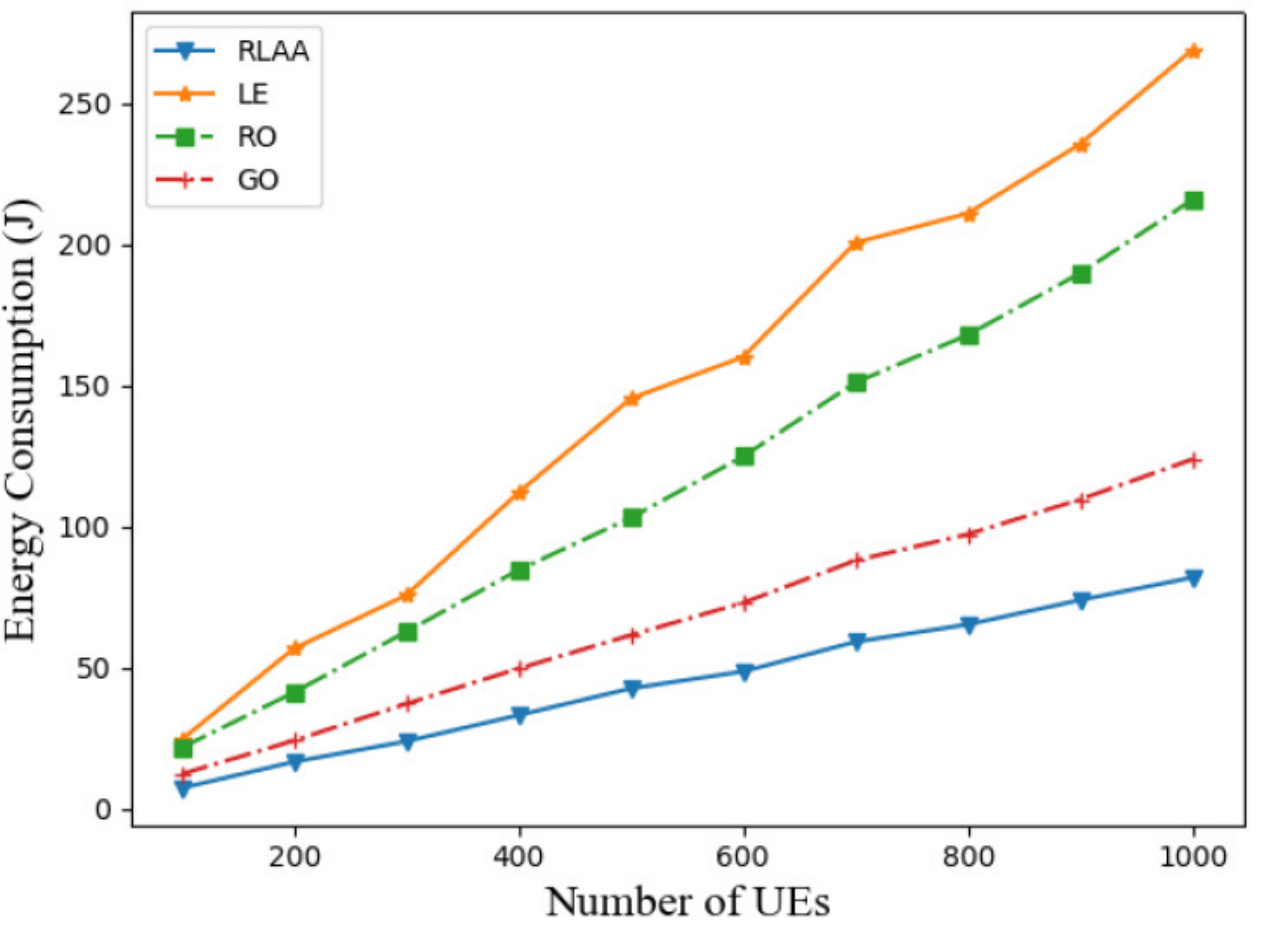}
	\caption{The overall energy consumption of RLAA, LE, RO and GO versus the number of UEs with 5 UAVs.}\label{fig4}
\end{figure}

In Fig.~\ref{fig4}, we further increase the number of UAVs to 5, where the center coordinates are set to as $[1200, 1200, 350]$, $[-1200, -1200, 350]$, $[-1200, 1200, 350]$, $[1200, -1200, 350]$ and $[0, 0, 350]$, respectively. One sees that our proposed RLAA still outperforms other compared algorithms, with significant amount of energy being saved for all the UEs.

\section{Conclusion}
In this paper, we studied a multi-UAV enabled MEC system, in which the UAVs are assumed to fly in circles over the ground UEs to provide the computation services. The proposed problem is formulated as a MINLP, which is hard to deal with in general. We propose a RLAA algorithm to address it effectively. Simulation results show that RLAA can achieve the same performance as the exhaustive search in small scale cases, whereas in large case scenario, RLAA still have considerate performance gain over other traditional approaches.

\section{Acknowledgements}
This work was supported in part by the Zhongshan City Team Project (Grant No. 180809162197874), National Natural Science Foundation of China (Grant No. 61620106011 and 61572389) and UK EPSRC NIRVANA project (Grant No. EP/L026031/1).

\bibliographystyle{ieeetran}
\bibliography{uav-collection}
\end{document}